# Influence of edge Laser-Induced Periodic Surface Structures (LIPSS) on the electrical properties of fs laser-machined ITO microcircuits

A. Frechilla[1], E. Martínez[1], J. del Moral[2], C. López-Santos[2,3], J. Frechilla[1], F. Nuñez-Gálvez[2,3], V. López-Flores[2,3], G.F. de la Fuente[1], D. Hülagü[4], J. Bonse[4], A. R. González-Elipe[2], A. Borrás[2], L.A. Angurel[1]

1. Instituto de Nanociencia y Materiales de Aragón, INMA, CSIC-Universidad de Zaragoza, María de Luna, 3, 50018 Zaragoza, Spain

2. Nanotechnology on Surfaces and Plasma, Instituto de Ciencia Materiales de Sevilla, ICMS, CSIC-Universidad de Sevilla, Américo Vespucio 49, 41092 Sevilla, Spain

3. Dpto. Física Aplicada I. Escuela Politécnica Superior. Universidad de Sevilla. c/ Virgen de África 7, 41011 Sevilla, Spain

4. Bundesanstalt für Materialforschung und –prüfung (BAM), Unter den Eichen 87, 12205 Berlin, Germany

**Abstract**
Scalable and cost-effective methods for processing transparent electrodes at the microscale are transversal for advancing in electrochemistry, optoelectronics, microfluidics, and energy harvesting. In these fields, the precise fabrication of micrometric circuits and patterns plays a critical role in determining device performance, material compatibility, and integration with added-value substrates. In this context, Laser Subtractive Manufacturing stands out among microfabrication techniques for its adaptability to diverse materials and complex configurations, as well as its straightforward scalability, affordability, and eco-friendly nature. However, a challenge in micromachining metals and metal oxides is the inherent formation of Laser-Induced Periodic Surface Structures (LIPSS), which can significantly impair electrical conductivity, particularly when circuit dimensions fall within the micrometer range. Herein, we investigate the micromachining of transparent conductive oxides (TCOs) using ultrashort pulse laser systems applied to indium tin oxide (ITO) thin films. We analyze the formation of LIPSS at the edges of the micromachined regions associated with the Gaussian distribution of the energy within the laser spot and their impact on the electrical properties depending on the circuit characteristics. Thus, we systematically evaluate the influence of LIPSS orientation and periodicity by fabricating various circuit patterns using femtosecond lasers at green (515 nm) and ultraviolet (UV) (343 nm) wavelengths. A correlation between electrical resistivity measurements and structural analysis, as determined by FESEM and TEM imaging, along with confocal topography, reveals distinct effects of nanostructure formation depending on the laser source. For green wavelength, the regions where LIPSS are oriented perpendicular to the ITO track exhibit higher resistance, by a factor just above two, compared to those where LIPSS are parallel. Additionally, UV laser processing results in a pronounced reduction of ITO thickness at the boundary between the LIPSS region and the substrate. The mechanisms for the formation of LIPSS with both wavelengths are also discussed. Furthermore, in narrow conductive tracks ranging from 6 to 8 µm, the impact of LIPSS is particularly significant, as the structured region occupies a dominant fraction of the total width.

# 1.- Introduction

Indium Tin Oxide (ITO), typically ≈90% wt% $In_2O_3$ + 10 wt% $SnO_2$, is a captivating material due to its low electrical resistivity and high transparency within the visible spectrum. These properties make it one of the most widely used Transparent Conducting Oxides (TCOs) in a variety of advanced applications [1]. ITO is particularly important for the fabrication of transparent electronic circuits [2] and serves as a universal electrode in various devices [3]. For example, ITO is employed in the manufacturing of Liquid-Crystal Displays (LCDs) [3,4], Organic Light Emitting Diodes (OLEDs) [5,6], or solar cells [7,8], and constitutes a promising material in sensing applications [9,10]. These technological applications led to the consideration of indium as a critical raw material in the European Union over the past few decades. However, this status has recently changed due to the emergence of new domestic suppliers [11].

The electrical and optical properties of ITO films are highly dependent on their fabrication procedure [12], which directly affects their performance for most applications [5,13]. For many advanced electrical applications, high-resolution patterning of thin films in the form of interconnected lines, paths, or assemblies [7] with characteristic features in the micrometric range constitutes the basis of an ample range of circuit designs. Thus, downsizing the conductive paths for transparent electrodes has paved the way for the development of micro-LEDs [14], high-resolution smart screens [15], biosensors [9], digital microfluidics [16], lab-on-chip platforms [17], and transparent thin-film transistors [18], among others. As a result, various techniques are currently applied to pattern ITO electrode structures with well-defined edges and electrically insulating paths between the conductive tracks. Widely used techniques include photolithography and micro-contact printing procedures. However, these methods are complex, time-consuming, and require multiple processing steps, expensive equipment, and toxic chemicals. These drawbacks underscore the urgent need for direct non-lithographic patterning strategies for the fabrication of micrometric and well-defined circuit structures using ITO thin films as starting material.

In this context, ultrashort pulse laser micromachining has emerged as a paradigm-shifting technology, offering unprecedented precision, control, and versatility for the fabrication of electrical circuits from thin films of conductive and non-conductive materials [19,20]. This laser processing technique enables the selective removal of the ITO thin film from certain zones, thereby creating the intended ITO patterns [21]. This method opens new possibilities for fabricating intricate circuit designs, overcoming traditional limitations associated with conventional manufacturing processes. Furthermore, laser micromachining of ITO thin films offers high flexibility for circuit patterning designs and fast processing times, making it especially attractive for large-scale fabrication. Additionally, laser patterning, as a mask-free technique conducted under ambient conditions, can be easily implemented for repairing applications [12].

Several studies have investigated the ablation process associated with the laser micromachining of ITO circuits using different laser sources. Some have reported the application of nanosecond pulsed lasers operating in the ultraviolet (250–350 nm) [2,10] or near-infrared wavelengths (around 1064 nm) [10]. Notably, 355 nm laser radiation has shown promising results for ITO ablation, likely due to the good matching between the photon energy (3.49 eV) and the material bandgap (3.6–3.8 eV) [10] enabling linear energy absorption in the material. Other studies have focused on the use of ultrashort laser pulses with durations in the femtosecond or picosecond range, using lasers with wavelengths between 700 and 800 nm [5,7,8,12,22] or 1030 nm [13]. While thermal effects may degrade patterning precision with longer pulse durations [10], ultrafast laser ablation has gained increasing interest due to its enhanced versatility and minimal detrimental thermal effects [23,24] and the possibility to process glass supported films in superstrate geometry and minimal device damage [9, 10, 13, 25]. In some cases, femtosecond laser patterning has been also complemented with a chemical etching process to complete the removal of the non-irradiated ITO layer [8,22].

One of the effects observed during direct ultrafast laser patterning of ITO thin films is the formation of Laser-Induced Periodic Surface Structures (LIPSS), which emerge due to the interaction between the laser beam and the material surface [26–29] These structures can significantly influence the optical and electrical properties of the patterned circuits, leading to anisotropic surface characteristics that can either enhance or degrade the performance of the final device [29]. LIPSS formation has been widely reported in different materials processed with femtosecond lasers, showing a strong dependence on the laser wavelength, fluence, number of pulses, and material properties [29–34]. However, their presence and effect in laser micromachined ITO structures, particularly at the edges of conductive tracks for micron size electrodes, so far remained an open topic requiring further investigation.

In this article, femtosecond Laser Subtractive Manufacturing (LSM) was used to fabricate micro-scale ITO conductive surface patterns. This technique utilizes a laser beam to selectively ablate specific areas of the ITO thin film, creating the "negative" (electrically isolating) space required for the circuit design. The remaining ITO material defines the desired conductive circuit pattern on the substrate. Specifically, we focus on the analysis of the LIPSS formed at the edges of the ITO tracks, due to the Gaussian energy distribution of the ultrashort laser pulses. To gain a deeper understanding of the topography, nanostructure characteristics and chemical composition of these transitional zones, we employ multiple characterization techniques, including Field Emission Scanning Electron Microscopy (FESEM), Transmission Electron Microscopy (TEM), electron probe microanalyzer and confocal microscopy topography measurements. Additionally, we assess their impact on the electrical behavior of the patterned circuits based on a model of parallel resistance networks in combination with four-probe resistivity measurements at macro and microscales. This study provides new insights into the role of LIPSS in femtosecond laser-processed ITO films and offers valuable guidelines for optimizing the laser patterning process to enhance device performance.

## 2.- Experimental

Fabrication of the electrical circuits was performed by selective laser irradiation on a commercial ITO thin film, ≈115, 140 or 400 (±5) nm-thick, deposited on a sodalime glass 2.5 cm x 2.5 cm substrate (supplied by Xop Glass).

### 2.1 Morphology, chemical and electrical characterization methods

Surface morphology was analyzed with a FESEM (MERLIN Carl Zeiss, Oberkochen, Germany) using secondary electron (SE), backscattered electron (BE) and in-lens detectors. The electron beam acceleration voltage was set to 5 kV enabling a high surface sensitivity. Surface topography was also characterized using confocal microscopy (2300 Plμ Sensofar, Spain). Image post-processing and 2-dimension fast Fourier transforms (2D-FFT) were carried out using the open-source software Gwyddion (vers. 2.61, Czech Metrology Intitute, Brno, Czech Republic). To analyze the topography of cross sectional areas near the laser micromachined edges, some lamellas were prepared using a Focused Ion Beam (FIB) in a Dual Beam Helios 650 (FEI company, Hillsboro, OR, USA). Scanning transmission electron microscopy (STEM) imaging of these lamellas was performed in a probe-corrected Titan (Thermo Scientific) operated at 300 kV and equipped with a high brightness X-FEG and a spherical aberration Cs-corrector (CEOS) so that the condenser system yields a sub-angstrom probe size. High angle annular dark field (HAADF) images were obtained with a HAADF detector by Fischione (Pittsburgh, PA, USA).

Chemical surface characterization was performed by a JXA-iHP200F (JEOL Ltd., Akishima, Japan) electron probe microanalyser, which is based on FESEM and is equipped with a SS-94040SXSER spectrometer. Wavelength-dispersive X-ray spectroscopy (WDS) was used for high spectral resolution, in order to detect possible compositional variations of the ITO thin films near the laser micromachined paths. Specifically, the elements In, Sn and O were analyzed, working at 6 kV and 10 nA electron acceleration voltages for excitation.

The electrical properties were measured with a Keithley 2000 multimeter connected to a 4-point probe station. The measurement configuration was performed with a distance between voltage contacts of 4.3 mm. All the measurements were performed at air atmosphere and at room temperature. Selected samples were characterized using a FESEM Zeiss GeminiSEM 300 at 5 kV with a Kleindiek micromanipulator system in a 4-point probe configuration assisted by a Keithley 2635 multimeter source, using a set of tungsten Micro-Pico Probes T4-10 of 3.3 mm length, 10 µm diameter, and ~4 ± 0.1 µm point radius with a distance between voltage contacts of 60 µm (a micrograph of the four probes emplaced on an ITO conductive track is shown in Figure S1 of Supplementary Information).

## 2.2 Laser machining procedure

A femtosecond laser system (Carbide CB3-40W+CBM03-2H-3H, Light Conversion, Lithuania) operated at two different wavelengths $\lambda$ = 515 nm—green or 343 nm—UV (corresponding to the second and third harmonic, respectively) was used to remove the ITO layer in preselected areas. At the working distance, the beam has elliptical Gaussian profiles with $1/e^2$ intensity decay major axes of ≈ 50 µm (green) and 64 µm (UV), with the main axes ratio of 0.97 and 0.89, respectively. For the experiments, the laser beam was linearly polarized and the pulse duration was 249 fs (green) or 238 fs (UV). All irradiations were performed in air using a substrate geometry, i.e. with the film side facing the focused laser beam [35].

In this work, several ITO tracks, of widths varying from ≈ 10 to 1000 µm were generated using LSM. The ITO layer was removed in the predefined areas using the laser beam scan method, which employed a galvanometric mirror system along with a f-theta lens of 330 mm focal length, controlled by dedicated software (Direct Machining Control, UAB, Lithuania). Following the procedure described previously [32] to achieve homogeneous energy distribution across the thin film, we obtained the machining parameters that were used in this work: repetition frequency of 10 kHz, scan speed of 25 mm/s, hatching distance between consecutive scanning lines of 5 µm and pulse energies of 6.55 µJ ($\lambda$ = 515 nm) and 3.40 µJ ($\lambda$ = 343 nm). The laser scan direction was always parallel to the ITO track and for each wavelength two cases were analyzed: with the laser polarization either parallel or perpendicular to the track.

## 3.- Results and discussion

Initially, the ITO ablation thresholds for both wavelengths were determined following the method described by Liu [36]. A multipulse analysis was then performed, varying the energy per pulse ($E_p$) and the number of pulses ($N$). As an illustrative example of these experiments, Figure 1 shows the spots produced in a film irradiated with 1 or 10 pulses and $E_p$ = 6.55 µJ ($\lambda$ = 515 nm) or 3.40 µJ ($\lambda$ = 343 nm). It is noted that when using a wavelength of 515 nm, the laser-marked region is nearly circular, with a diameter of approximately 16 µm for $N$ = 1. In contrast, when working with 343 nm wavelength, the marked region has an elliptical shape, with short and long axis sizes of about 22 µm and 25 µm, respectively. Also, these images show the higher level of absorption of the UV radiation in comparison with the visible one. At $N$=10 and for both wavelengths, we obtained a central area where the ITO layer was removed entirely (black contrast in the images), with a non-steep boundary between the ablated and non-ablated areas. This transition is caused by the Gaussian energy distribution of the laser pulse. In this case, the onset of complete ITO removal was found to occur at pulse fluences of approximately 32.6 J·cm$^{-2}$ for 515 nm and 7.9 J·cm$^{-2}$ for 343 nm, with 10 pulses ($N$ = 10). Thus, as the local pulse fluence values evolve continuously from the ablation threshold to zero, there is a region between the ablated and the pristine film where the laser has modified the original surface, even forming some spatial periodic nanostructures (see zoomed images in the figure). The topographic, chemical and electrical characteristics of these transitional zones, as well as their extent, should be considered when designing ITO microcircuits. Figures also show the strong differences observed in the characteristics of these regions depending on the selected wavelength.

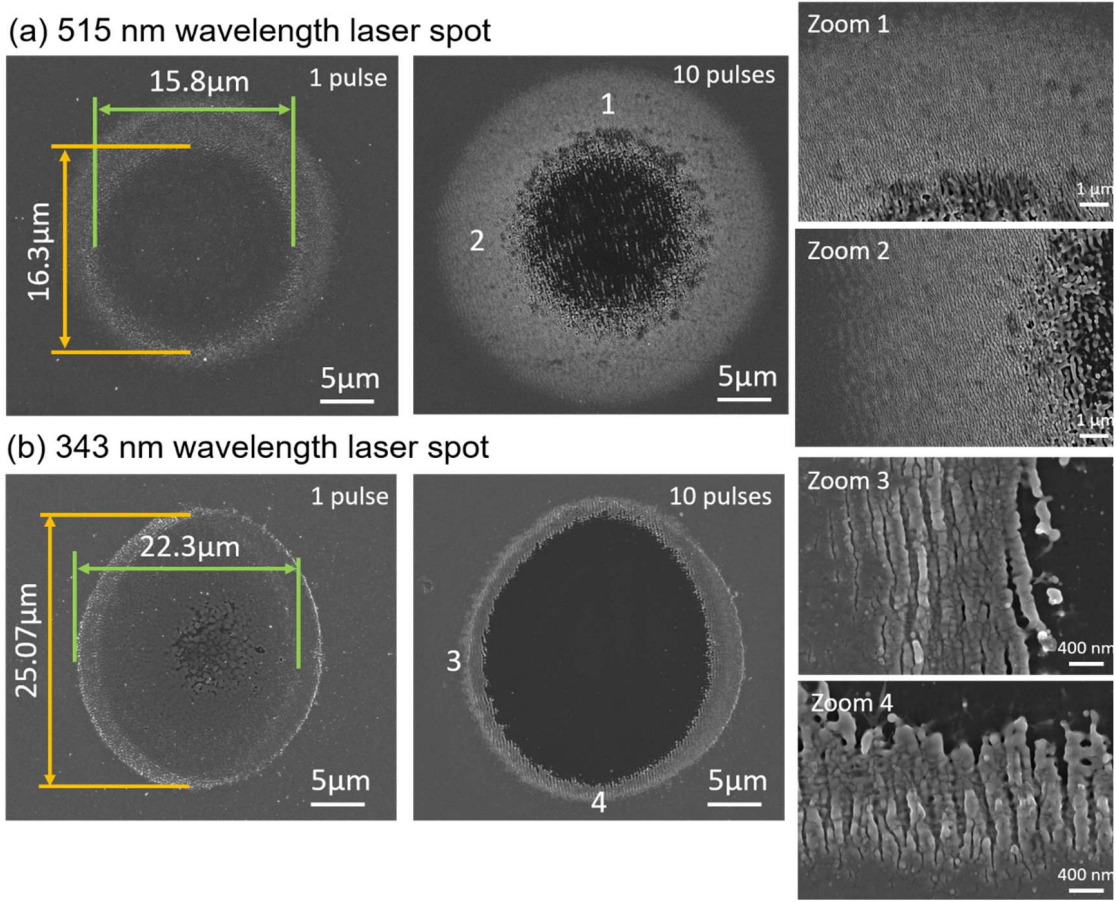

*Figure 1.* FESEM micrographs (in lens detector) of the sample surface after having applied 1 or 10 laser pulses in a given position with (a) λ = 515 nm radiation and $E_p$ = 6.55 µJ and (b) λ = 343 nm and $E_p$ = 3.40 µJ. The substrate appears in black contrast in the images. The laser beam polarization direction lays almost parallel to the horizontal direction in all these images.

### 3.1. Nanostructures at the transition zones between the machined and non-machined regions using λ = 515 nm

Figure 2 shows SEM images of the transitional area between the ITO track (left) and the glass substrate (right) when machining with λ = 515 nm for both laser polarization configurations. Four differentiated areas can be identified in these images, highlighted by distinct colour bar segments at the top of the low magnification images: a region where the original ITO has not been modified by the laser treatment (**red**), a laser nanotextured region with a dense LIPSS structure (**green**), a region of isolated LIPSS (**orange**) leading to a loss of film continuity, and finally the underlying glass substrate (**blue**), where the covering ITO layer was completely removed by the laser treatment. Focusing on the transitional zones it is noteworthy that in both configurations, i.e. LIPSS parallel or perpendicular to the electrically conductive ITO track, LIPSS gradually emerge and become increasingly well-defined as the distance from the pristine zones of the ITO film increases.

The LIPSS ridge orientation is defined here by the direction of the laser beam polarization, which is perpendicular to the direction of the electric field vector of the laser radiation. Towards the border of the non-modified ITO, the spatial periodicities are below 100 nm far below the laser irradiation wavelength. Thus, these LIPSS can be classified here as high spatial frequency LIPSS (HSFL, type I) formed on dielectrics [37] and that must emerge here on the ITO film through a collective near-field optical scattering of surface defects mechanism [38].

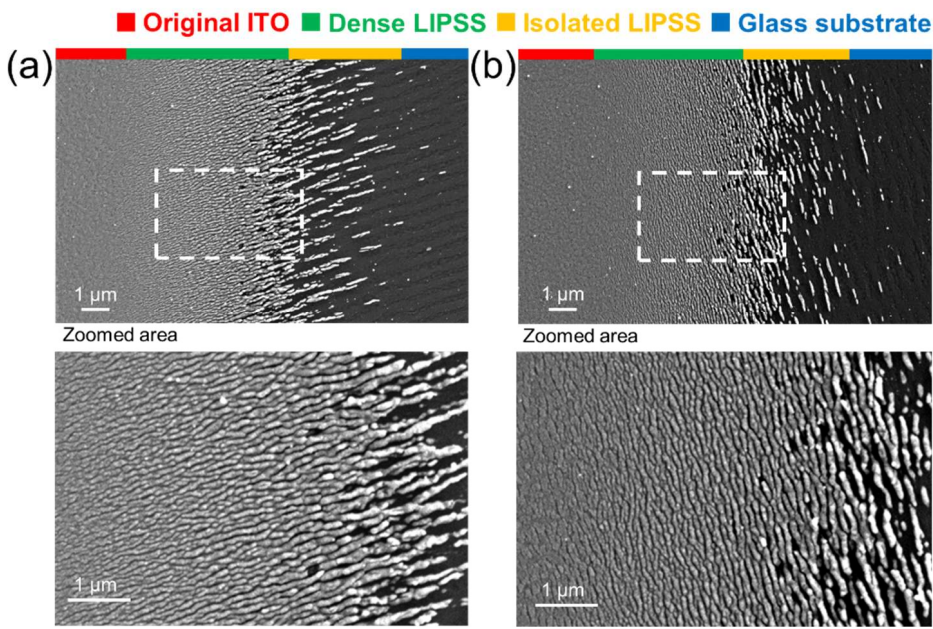

*Figure 2. Top-view FESEM (SE) images of the edge of an ITO track micro-machined with 515 nm fs-laser wavelength: (a) LIPSS perpendicular to the electric track. (b) LIPSS parallel to the track. The original ITO and the glass substrate appear on the left- and right-hand sides of the micrographs, respectively. The top row panels provide overview images, while the bottom row ones display higher magnification images of the areas marked by a dashed white box.*

The evolution of the LIPSS spatial periodicity in the transitional region is studied by 2D-FFT analysis of several regions at the boundary between the ITO track and the glass, as shown in Figure 3(a). Thus, from a FESEM image with approximate dimensions $x \times y$ = 22 µm × 15 µm around the ITO track's edge (directions perpendicular × parallel to the track), several image regions of size $\Delta x \times \Delta y$ ($\Delta x$ = 1 µm in most cases and $\Delta y$ = 15 µm) were selected, analyzing the 2D-FFT maps of all of them. The corresponding central 1D-profiles perpendicular to the HSFL ridge direction are shown in Figure 3 (b). This will allow obtaining the spatial periodicity in each position and estimating the variation of the spatial periodicity in the transitional zone, which is displayed in Figure 3 (c). Figure S2 of the Supplementary Information shows more details of the performed analysis.

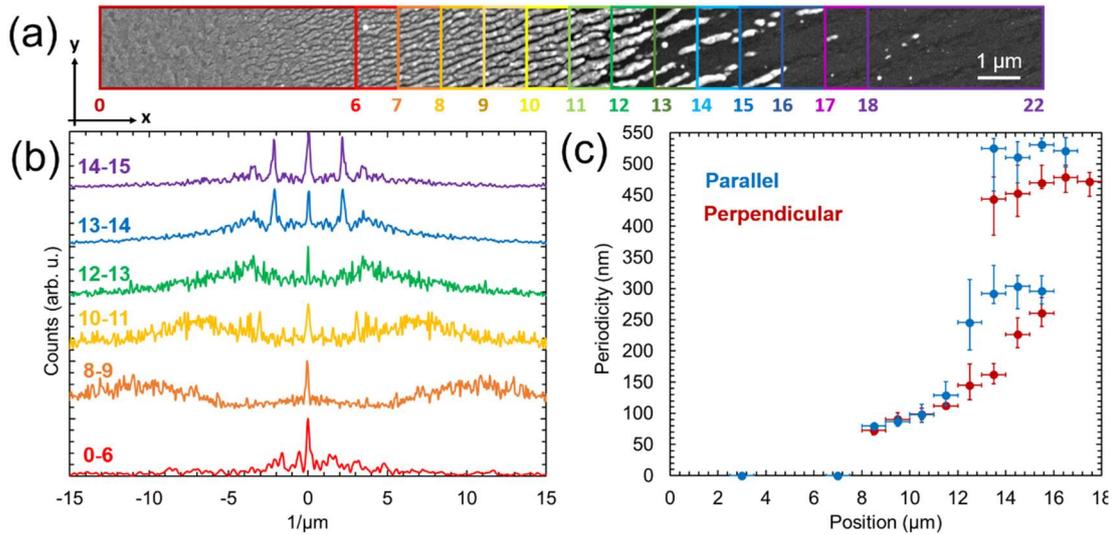

*Figure 3. Analysis of the spatial periodicity evolution along the ITO/glass boundary for λ = 515 nm: Top-view SEM image (a) showing the position ranges of the x-coordinate (in μm) that define the different sections analysed by 2D-FFT. (b) Central 1D-profiles perpendicular to the HSFL ridge direction of the 2D-FFT images, used to quantify the mean spatial periodicity of the LIPSS in different sectors of the transitional zones (c). Note that (c) also shows the results for LIPSS parallel to the ITO track, for comparison.*

These analyses reveal a clear trend in the evolution of the surface patterns. In more detail, initially, i.e. closer to original ITO, a high spatial-frequency structure is observed, with periodicities ranging from 90 to 100 nm. This fine nanostructuring likely corresponds to the early-stage of ablation with LIPSS formation, where plasmonic coupling at transient metallic defects and collective interference effects begin to modulate the surface at subwavelength scales in the form of HSFL-I. When approaching the glass substrate region (i.e. upon increasing the coordinate *x* in the figure), where the local laser fluence is increased due to the Gaussian laser beam profile, the LIPSS periodicity gradually increases, eventually reaching a regime where two distinct spatial frequencies coexist. Specifically, a dominant well-defined spatial periodicity is observed at ≈ 450 – 470 nm, which would correspond to low spatial frequency LIPSS (LSFL, type I), typically expected on absorbing materials at about 0.8 – 0.9 times the laser wavelength on absorbing materials [29,33].

This transition from HSFL-I characteristics over a dual-period region towards a LSFL-I zone suggests a dynamic evolution of the surface interaction mechanism with the laser scanned beam depending on fluence: in the low fluence wing of the Gaussian beam distribution, HSFL-I LIPSS are seeded and formed at the surface of the ITO film via the above mentioned collective optical near-field scattering at localized individual plasmonic defects. In the high intensity (center) regions of the scanned laser beam the combination with multi-pulse incubation effects leads to LSFL-I LIPSS seeding and formation at the early stage of ablation, originating under the action of the well-known Surface Plasmon Polariton (SPP) mediated scattering and interference mechanism [37]. Due to the larger local fluence, in this zone, the ITO film is gradually removed upon excitation by several laser pulses, leaving behind a bare glass substrate, where a minimal damage is only induced at the interference maxima of the LSFL-I pattern. In the transitional (intermediate) fluence range, spatially separated HSFL-I on ITO can survive at the interference minima of the initially formed LSFL-I LIPSS.

A cross-sectional view of the nanostructured ITO film obtained by TEM is displayed in Figure 4. The image corresponds to an ITO track edge machined with 515 nm wavelength, using a polarization/scanning configuration with LIPSS aligned parallel to the track, and the lamella was taken perpendicular to them. The left-hand side of the overview image provided in the top panel corresponds to the pristine ITO film deposited on the glass substrate. Towards the right-hand side direction, we first notice some small indents in the film (see area at position 1) of ~20 nm width that gradually become more pronounced and deeper (detailed views at positions 2 and 3), resulting in smooth deep trenches with a large depth-to-width aspect ratio ($A \gg 1$). Eventually, some additional shallower and wider near-surface surface depressions may appear in the ITO film (right-hand side in the top image). It is important to note that despite the surface structuring, the height of the ITO nanostructures (e.g. LIPSS) remains essentially unchanged throughout the entire textured region analyzed in this figure, coinciding with the thickness of the original ITO film. The periodic surface structures continue to develop into depth at increasing local fluences until they reach the glass substrate, spanning (laterally) a width of several micrometers ($\approx 4 - 6$ μm) covered by LIPSS generated on the ITO film. The observed large aspect ratio $A \gg 1$ of the nanostructures in region 3 further confirms that HSFL-I are formed in this region [37].

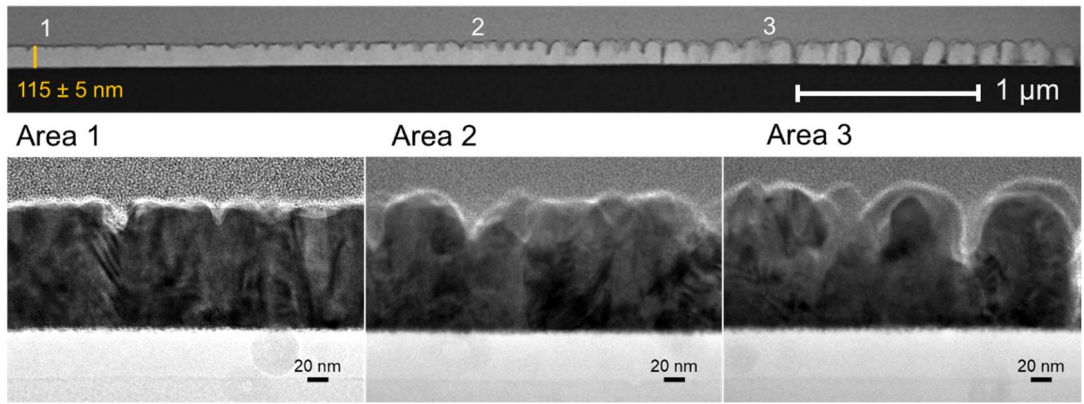

*Figure 4.* TEM cross-sectional views of the nanotextured ITO layer at a lateral position near the unaffected pristine film (left) upon machining with 515 nm laser wavelength (HSFL-I parallel to the ITO track): (Top) Low-magnification HAADF-STEM. The vertical orange line indicates the thickness of the pristine ITO layer. (Bottom) HRTEM images taken at higher magnification at the indicated positions (1 – 3).

### 3.2. Nanostructures in the transition zone between machined and non-machined regions using λ = 343 nm

Figure 5 displays the ITO/glass boundaries when machining with a laser wavelength of 343 nm, for both laser polarization orientations. As expected, LIPSS are also generated at the transition zone between the machined paths and the unmodified ITO film, depicting the four differentiated regions already mentioned. These regions are indicated in the figure with colour bar segments. However, notable differences are observed with respect to the 515 nm machining. On the one hand, following the differences observed in Figure 1, the nanostructured region is narrower with transitions between the different sectors considerably sharper. Note that the width of the isolated-LIPSS zone is well below 1 μm in this case. The "dense LIPSS" zone is also notably more compact and less extended, suggesting a more abrupt interaction and localized energy deposition with this wavelength. It is also noteworthy that the texturing is not completely homogeneous along the ITO track, since regions with better defined LIPSS structures coexist with areas where the ripples are clearly less developed. 2D-FFT analysis of the dense LIPSS regions gave spatial period for the nanostructures around 300 nm, as expected for LSFL-I. Interestingly, HSFL-I are not observed in this case. This is fully consistent with the general observation that HSFL-I structures form only for sub-band gap excitation by ultrashort pulses [37], i.e. when the material is transparent and the photon energy of laser is smaller than the band gap. While the latter condition is fulfilled at □ = 515 nm, at a wavelength of 343 nm, the ITO film absorbs the laser radiation in a relatively linear manner, allowing for the direct promotion of carriers into the conduction band through the absorption of a single photon. Thus, the UV wavelength directly promotes the formation of LSFL-I LIPSS via an ultrashort-pulse enabled global transient "SPP-activity" rather than via a nonlinearly-excited defect mediated local-plasmonic scattering with collective interference for the HSFL-I at 515 nm laser wavelength. Other periodic structures with periods ranging from 50 to 80 nm and oriented perpendicular to them were also observed between the LSFL features, in agreement with the observations reported in [39,40].

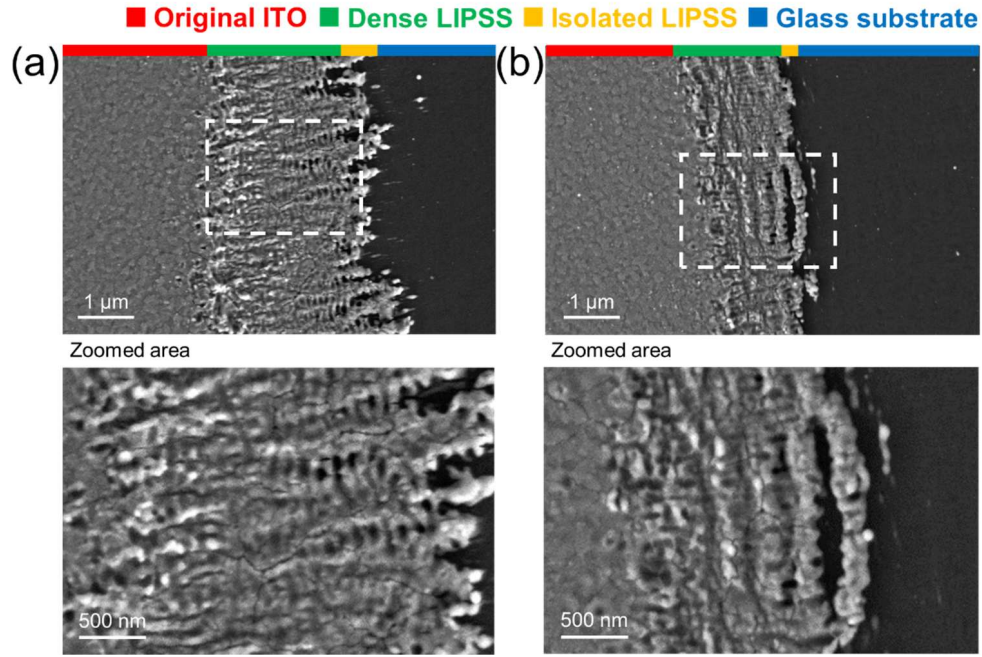

*Figure 5.* Top-view FESEM (SE) images of the boundaries between the ITO track and the glass after having applied the laser micromachining treatment with λ = 343 nm: (a) LIPSS perpendicular to the path. (b) LIPSS parallel to the path. The original ITO surface appears on the left-hand side of the micrographs, and the glass substrate (black contrast) on the right-hand one. The top row provides overview images, while the bottom row displays higher magnification images of the areas marked by a dashed white box.

Figure 6 displays cross-sectional views, obtained by TEM, of the ITO/glass boundary developed when machining with the 343 nm laser radiation. In this case, the laser-affected region exhibits a notable reduction in the ITO thickness, down to nearly 50% of the original value of ~140 nm of the pristine ITO film. This ablative thinning process continues progressively from the ITO film surface into its depth until the ITO layer completely vanishes at a distance of approximately 2.7 µm from the starting of the ablated region. Moreover, the high-aspect-ratio LIPSS trenches are not seen in this specific lamella, in agreement with the SEM observations previously shown in Figure 5. However, in contrast to the TEM image provided in Figure 4, the laser-induced surface corrugations exhibit a depth-to-width aspect ratio $A < 1$, as it can be expected for LSFL-I structures [37]. This feature further supports the above-given assignment for the LIPSS discussed when dealing with the 515 nm laser ablation.

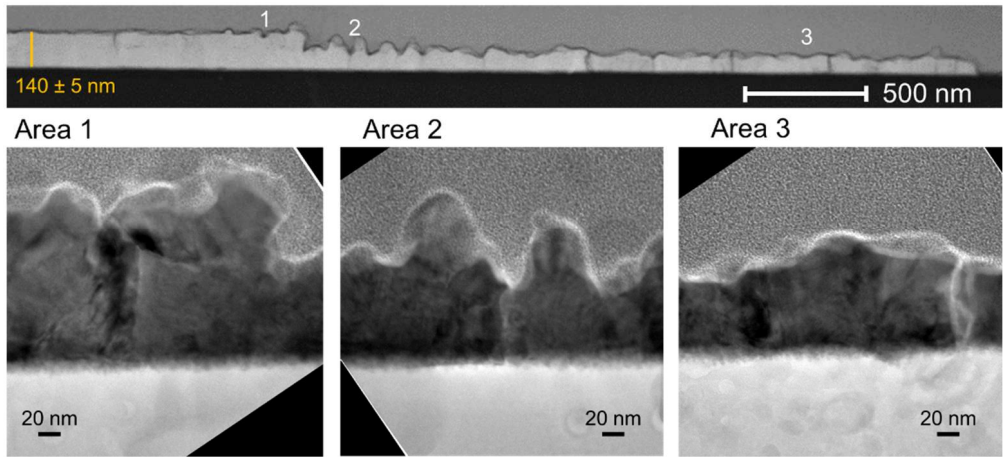

*Figure 6.* TEM cross-sectional views of the nanotextured ITO layer at a lateral position near the unaffected pristine film (left) upon machining with 343 nm fs-laser wavelength (LIPSS parallel to the ITO path): (Top) Low-magnification HAADF-STEM. The vertical orange line indicates the thickness of the pristine ITO film. (Bottom) HRTEM images taken at higher magnification at the indicated positions.

The above top-view SEM and cross-sectional TEM results suggest that using this UV laser wavelength the formation of LIPSS on ITO is dominated by near-laser-wavelength-sized LSFL-I due to laser-induced material removal from the surface (ablation) rather than a material-conserving hydrodynamic displacement of the melt with redistribution of material at and to above the initial surface plane. Nevertheless, the morphology of the observed small-scale spherical surface heterogeneities is probably the result of rapidly solidified protuberances formed in the residual melt layer during ablative LIPSS formation, indicating the existence of accompanying transient hydrodynamic effects acting at a few-nanometre-scale [39].

### 3.3. Chemical analysis of the laser induced changes on the nanostructured region on the edge of the ITO tracks

WDS maps obtained by electron probe microanalysis were used to study the compositional properties of the laser-affected zones of the ITO film at the edges of the micromachined paths. It is important to recall here that this technique has considerably higher spectral resolution and sensitivity compared to energy dispersive X-ray spectroscopy (EDS). This feature is particularly relevant in this material due to the low Sn/In compositional ratio and the proximity of the X-ray emission lines for the elements Sn and In (for example, the $L_{\alpha 1}$ transition of Sn, 3.44 keV, almost overlaps with the $L_{\beta 1}$ transition of In, 3.49 keV).

Figure 7 displays the results obtained when using either 515 nm or 343 nm laser wavelengths to machine the ITO film, both with LIPSS parallel and perpendicular to the ITO track. The original ITO and the glass substrate appear on the left- and right-hand sides of the figures, respectively. Starting from the non-affected ITO film side, a decrease in the atomic concentration of both elements, Sn and In is generally observed as the transition zones are approached. When using a wavelength of 515 nm, this decrease is smooth at the beginning, which for this particular area is about 12% in the region between $x = 3$ μm and 5.5 μm, coinciding with the appearance of high-frequency LIPSS, and becomes sharper (up to a 62%) as it approaches the isolated-LIPSS region (marked by an orange bar at the top of the graphs). The complete lateral transition for 343 nm laser wavelength is considerably narrower than for 515 nm (2.6 μm vs. 8.0 μm, respectively, in these analyzed zones).

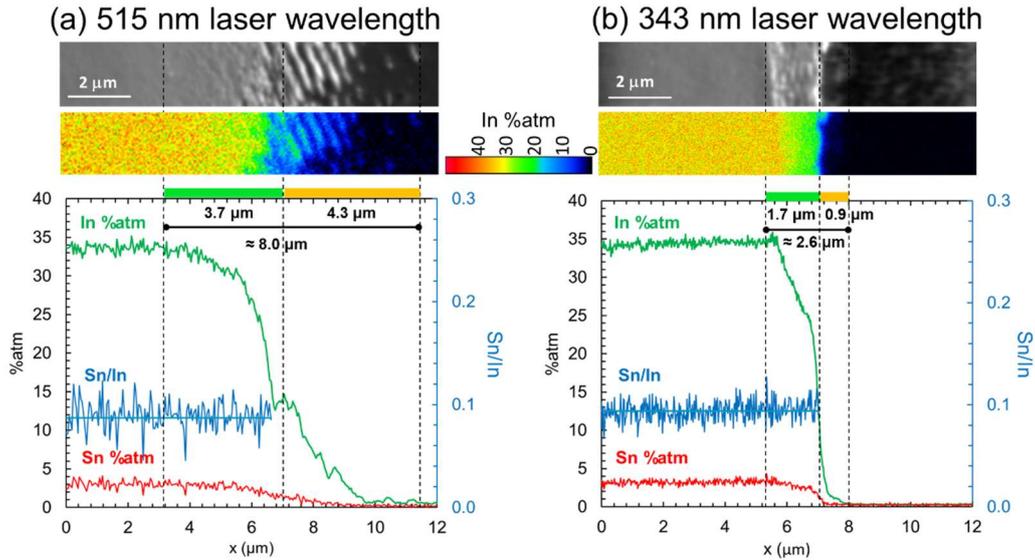

*Figure 7.* WDS microanalysis of the ITO films near the micromachined paths using: (a) λ = 515 nm and (b) λ = 343 nm laser wavelengths. The original ITO and the glass substrate appear in the left- and right-hand sides of the images and graphs, respectively. LIPSS are parallel to the ITO paths in both cases. Top row panels: SEM (SE) images of the analysed areas. Middle row panels: Corresponding WDS maps of In concentration (%atm) in the same areas. Bottom row panels: Line profiles of In, Sn atomic% concentration and Sn/In elemental ratio along both transitions using an averaging line width of 2 μm (i.e., each point in the graph is the average of the values in the map at a given x-position). The darker blue line corresponds to the average Sn/In ratios in this zone, 0.087 (0.015 std dev, green laser wavelength) and 0.094 (0.009 std dev, UV laser wavelength). Note that oxygen concentration was not included here for clarity purposes, but it can be seen in the supplementary file (figures S3 and S4).

In both cases no significant changes in the overall chemical composition of the ITO films are observed, consistently maintaining a constant Sn/In ≈ 0.09 atomic ratio in the transition, similar to the composition of the original ITO film. For example, in the case of green wavelength shown in the figure, in the region between $x = 0$ and 4 μm (original ITO film) the average value is 0.089 (0.015 std dev). At the beginning of the transition, 4 μm < $x$ < 6 μm, this ratio remains similar 0.090 (0.012). Finally, in the zone that exhibits a sharper decrease of In and Sn (6 μm < $x$ < 9 μm), this value increases slightly up to 0.096 (0.029). Note that the values of the Sn/In ratio for Sn concentration lower than ≈ 1.5 %atm were not included in the figure because of the increment of noise for the selected microanalysis parameters. In general, we can conclude that the observed Sn and In variations are in good correlation with the changes in the film topography at the track's edges seen in TEM images (Figures 4 and 6). Thus, the decrease of In and Sn elements would be directly linked to a relevant contribution of the substrate as the ITO layer becomes thinner (or holes start forming). This conclusion is further supported by the increase in the oxygen signal, which would correspond to the glass underneath the ITO, as seen in Figures S3 and S4. Moreover, the transition onset for UV is marked by a small but discernible increase of In and Sn (note the higher density of red dots in the map of Figure 7 (b) just at the original/transition area), followed by a sharper decrease of In and Sn. These analyses were performed in other areas obtaining similar results.

### 3.4. Influence of the laser-generated nanostructures on the electrical properties of ITO tracks

The nanostructured regions formed at the ITO glass transition zone (i.e. LIPSS zone) are expected to have some influence on the electrical properties of the laser micromachined tracks. This could be crucial in the cases where the width of the transitional zone becomes comparable to the ITO track's width. To quantify these effects for the four analysed configurations (i.e., 343 and 515 nm laser wavelengths, with LIPSS either parallel or perpendicular to the path), several ITO tracks with different widths ranging from 1000 µm down to about 20 µm were micromachined for each case. A picture of a 12-tracks set, each with its corresponding current and voltage contacts for four-point resistance measurements, is shown in Figure S5 of the supplementary data.

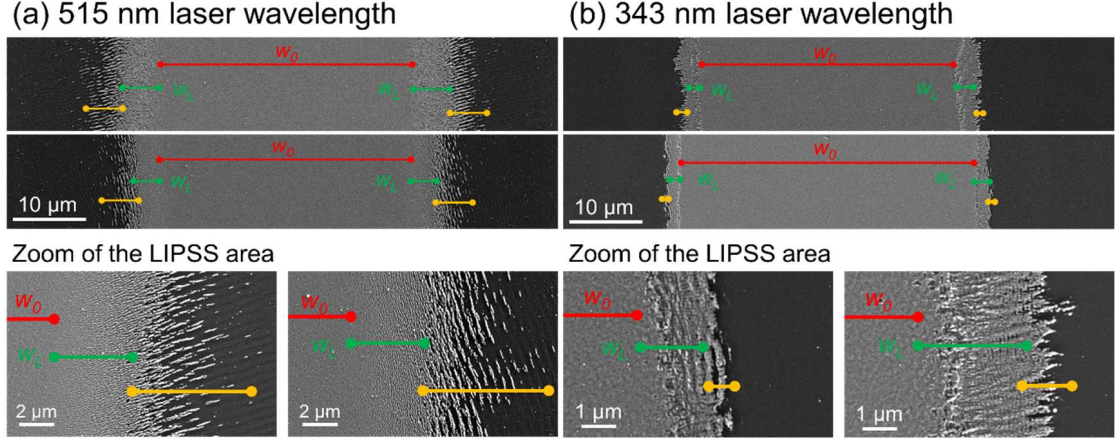

*Figure 8. FESEM (SE) images of four ITO tracks machined with different configurations: (a) 515 nm, LIPSS parallel and LIPSS perpendicular, (b) 343 nm, LIPSS parallel and LIPSS perpendicular. Colour segments correspond to the differentiated regions by their nanostructure characteristics: "original" (red), "dense LIPSS" (green) and "isolated LIPSS" (orange). The track width is around 40 – 45 µm for all configurations.*

Considering the results of the previous sections, the electrical resistance of the laser-generated ITO fingers has been theoretically modelled as several resistances in parallel. Thus, the central zone of the track, of width $w_0$, would correspond to the original ITO (red colour in Figures 2, 5 and 8). Its resistance, $R_0$, is assumed to retain both, the resistivity ($\rho_{ITO}$) and thickness ($t$) of the original ITO film, as this region was not affected by the laser irradiation. Both adjacent regions, with a width, $w_L$, coloured in green in Figures 2, 5 and 8, are associated with the presence of dense LIPSS. We assume that both regions on each side would have the same electrical resistance, $R_L$ (for symmetry) but different from that of the original ITO film. This electrical behaviour can be modelled with Equations (1) and (2).

$$R_0 = \rho_{ITO} \cdot \frac{l}{t \cdot w_0} \tag{1}$$

$$R_L = \rho_L \cdot \frac{l}{t_L \cdot w_L} = \alpha \cdot \left(\frac{\rho_{ITO}}{t}\right) \cdot \frac{l}{w_L} \tag{2}$$

The length between voltage contacts was $l$ = 4.3 mm in all cases. The proportionality factor $\alpha$ in Equation (2) accounts for differences compared with the original ITO in both, the resistivity ($\rho_L$) and the effective thickness ($t_L$) values, due to nano-structural modifications and to the reduction of the effective thickness produced by undulations and material ablated mechanisms [26].

The proposed model considers that the values of $R_L$ and $w_L$ are the same for all the measured tracks fabricated with a given laser wavelength and LIPSS orientation (i.e. $\alpha$, $w_L$ and $R_L$ are constant for each four analysed configurations). Finally, we assume that the external areas (orange colour in Figure 8), where LIPSS become spatially and electrically isolated, would not contribute to the total conductivity due to the loss of percolation paths. With the above hypotheses, the estimated track resistance, R, is given by Equation (3).

$$\frac{1}{R} = \frac{1}{R_0} + \frac{2}{R_L} = \frac{t}{\rho_{ITO} \cdot l} \cdot w_0 + \frac{t}{\rho_{ITO} \cdot l} \cdot \frac{1}{\alpha} \cdot 2w_L = \frac{1}{R_0}\left(1 + \frac{1}{\alpha} \cdot \frac{2w_L}{w_0}\right) \quad (3)$$

Figure 9 shows the ratio between $R_0$ and $R_{meas}$, for varying track widths, $w_{track} = w_0 + 2 \cdot w_L$, in the four analysed configurations. It is seen that these values follow the expected trend according to the proposed model, as shown in Equation (4).

$$\frac{R_0}{R_{meas}} = 1 + \frac{1}{\alpha} \cdot \frac{2w_L}{w_0} = 1 + \frac{1}{\alpha} \cdot \frac{2w_L}{w_{track} - 2w_L} \quad (4)$$

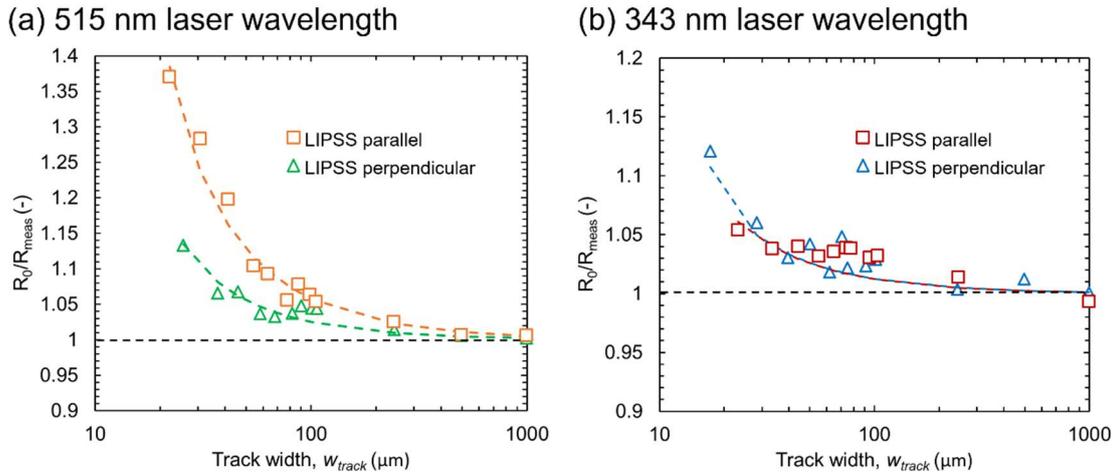

*Figure 9. Measured normalized electrical resistance ratio $R_0/R_{meas}$ of the ITO tracks as a function of the track's width $w_{track}$, plotted to estimate the contribution of the laser-affected edges to the total resistance of the ITO track for the four used configurations: (a) 515 nm, LIPSS parallel and LIPSS perpendicular, (b) 343 nm, LIPSS parallel and LIPSS perpendicular. The lines represent least-squares-fits to Equation (4), used to estimate the corresponding values of α, given in Table 1.*

Table 1 collects the α values estimated for each configuration from this analysis, together with the value $w_L$, which was previously determined by FESEM images (fixed for each configuration). The resistivity of the original ITO thin films $\rho_{ITO}$ was measured as $1.6 \cdot 10^{-6}$ Ω · m for the two batches, which is similar to the one reported in the literature for the same range of film thicknesses [1].

When machining using the green laser wavelength (Figure 9(a)), our results indicate that the nanostructured zones near the original ITO contributes significantly to the total resistance of the track, particularly for $w_{track}$ < 80–100 µm and LIPSS parallel to the track. Moreover, since the amount of ablated material seems similar for both orientations, the significant difference in the estimated α values (1.55 vs 3.5) would indicate an important electrical resistivity anisotropy of the laser-affected zone depending on the LIPSS orientation respect to the ITO track. As it can be expected for these HSFL-I LIPSS featuring isolating deep trenches into the bulk of the conductive film (A >> 1), the resistivity of the laser nanotextured region is higher when LIPSS are aligned perpendicular to the ITO track, by a factor just above two as compared with the configuration of parallel LIPSS.

On the other hand, when machining using UV wavelength (Figure 9(b) it was not possible to estimate with accuracy this anisotropy, probably because of the effect of the high amount of ablated material in this case, as discussed in previous Sections 3.1 and 3.2. Moreover, the surface corrugations of these LSFL-I exhibit a relatively shallow modulation ($A < 1$) and good lateral interconnection between LIPSS, a percolated morphology promoting an effective current passage across adjacent structures and mitigating orientation-dependent effects. From these considerations, it can be concluded that with this wavelength it is possible to neglect the contribution of the laser-affected regions to the total resistance $R$, within 5% accuracy, even for narrow conductive ITO tracks ($w_{\text{track}} >\approx$ 20–25 µm), as observed in the figure.

*Table 1. Values of the parameter α obtained from the proposed resistivity model. Average width of the nanotextured region, $w_L$, as estimated from FESEM images.*

| Laser wavelength and LIPSS orientation with the ITO tracks | $w_L$ (µm) | α |
|---|---|---|
| 515 nm, perpendicular | 4.1 ± 0.4 µm | 3.5 ± 0.5 |
| 515 nm, parallel | 4.1 ± 0.4 µm | 1.55 ± 0.05 |
| 343 nm, perpendicular | 2.0 ± 0.3 µm | 2.5 ± 0.5 |
| 343 nm, parallel | 1.8 ± 0.3 µm | 2.5 ± 0.5 |

One of the main objectives of this study was to minimize the influence of the generated LIPSS on the electrical conductivity of laser-processed ITO microcircuits. According to our results, this goal can be best achieved by using UV laser radiation with LIPSS (type LSFL-I) oriented parallel to the electrical track, i.e. by selecting a laser beam polarization perpendicular to it. Under these conditions the spatial extent of the region affected by the laser radiation is reduced (1.8 ± 0.3 µm), thereby preserving the conductive properties of the original film to a greater extent. On the other hand, the measured resistivity can be easily estimated with rather good accuracy (typically within 2–3%) using the proposed simple parallel circuit resistance model. However, the use of green wavelength to fabricate ITO paths of size $w_{\text{track}} > 20 – 30$ µm cannot be discarded because of the lowest α parameter (≈ 1.55) obtained using the green wavelength with LIPSS (type LSFL-I and HSFL-I) oriented parallel to the electrical path.

It is important to distinguish between the pitch, i.e., nominal distance between the laser tracks, and the effective width of the machined circuit defined by the laser-affected zone. This difference, which is more pronounced for the green wavelength because of its larger $w_L$, significantly influences the final electrical performance. In any case, the influence of the generated surface nanostructures on ITO becomes prominent for track widths narrower than 80 µm, as observed in Figure 9. Therefore, the proposed resistance model analysis is crucial for assessing the real electrical behaviour of the produced electrical circuits using a Laser Subtractive Manufacturing approach.

In order to explore the limits of our experimental approach, a series of conductive ITO tracks with widths of around 70–80 µm down to a few micrometres were machined for each configuration, and their resistance was measured "in-situ" in a SEM with a 4-point probe micromanipulators setup. Note that different sample pieces of ITO thin films with the same thicknesses (≈ 115 - 140 nm) were used for these tests. Figure 10 shows two representative FESEM images of these narrow tracks machined with green or UV wavelength in order to highlight the differences. Despite the similar widths along the length of the two lines, about 8–9 ☐m, they exhibit very different topographies. Thus, the one obtained with 515 nm laser wavelength is completely nanotextured by the laser, with well-defined LIPSS even at the centre. Conversely, the track micromachined with 343 nm wavelength has still preserved a central band of the original ITO film, ≈ 3.5 ± 0.5 µm in width. The plot of the measured electrical resistance as a function of the overall ITO track width (Figure 10c) reflects this feature as a large increase in the track resistance at around 10 µm, with an influence of the orientation of the LIPSS with respect to the current direction.

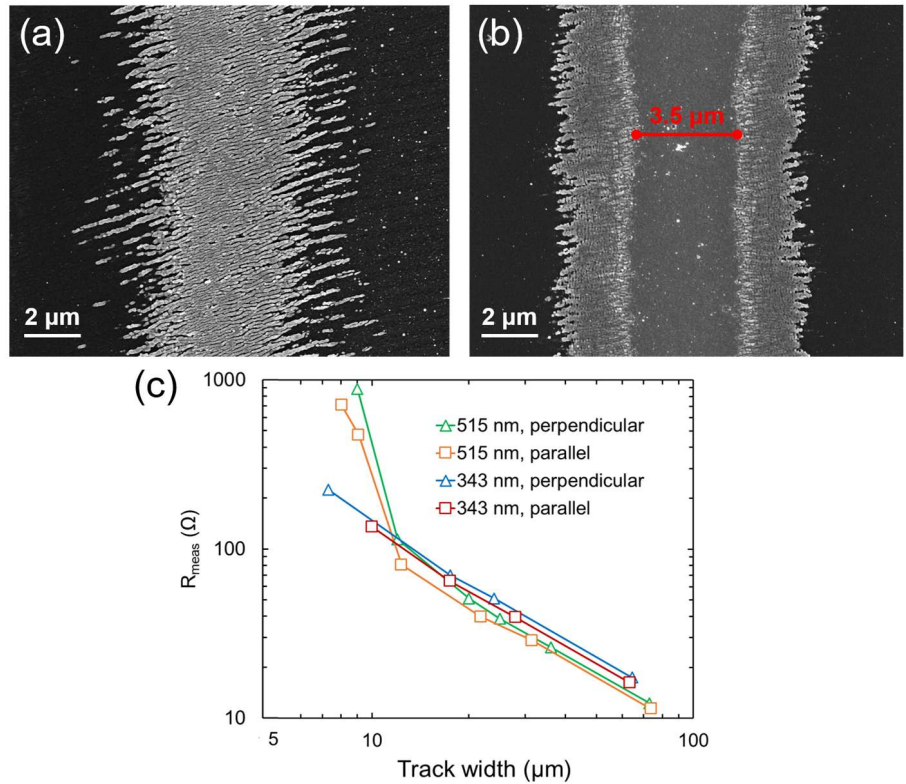

*Figure 10. Top-view FESEM (inlens) images of narrow ITO tracks (bright regions) micromachined with (a) 515 nm and b) 343 nm laser wavelengths, both with LIPSS perpendicular to the track. The segment in (b) marks the average width (3.5 µm) of the pristine ITO area. (c) Measured resistance for tracks of different widths, using a distance between voltage contacts of ≈ 60 µm.*

### *3.5. Fabrication of electrical circuits*

Based on the obtained results, several laser-machined circuits were fabricated as proofs of concept of this laser-based technology. Whether the objective is to optimize signal integrity in high-frequency applications or to maintain uniform resistance across conductive tracks, the ability to fine-tune and regulate the widths of both machined regions and remaining ITO lines in a contactless manner is paramount for achieving the desired circuit performance and reliability. For this purpose, multiple sets of parallel ITO lines, with different widths and distances between them, were machined. An example of these tests is presented in Figure 11. It shows nearly equidistant troughs with width of 35 µm machined on a 400 nm thick ITO film using the 343 nm fs-laser with LIPSS (LSFL-I) oriented parallel to the long paths. This 12-path circuit (with resistance 18.1 kΩ, very close to that expected considering the LIPSS formed at the edges) demonstrates a proper control over both the widths of the machined regions and the remaining ITO tracks.

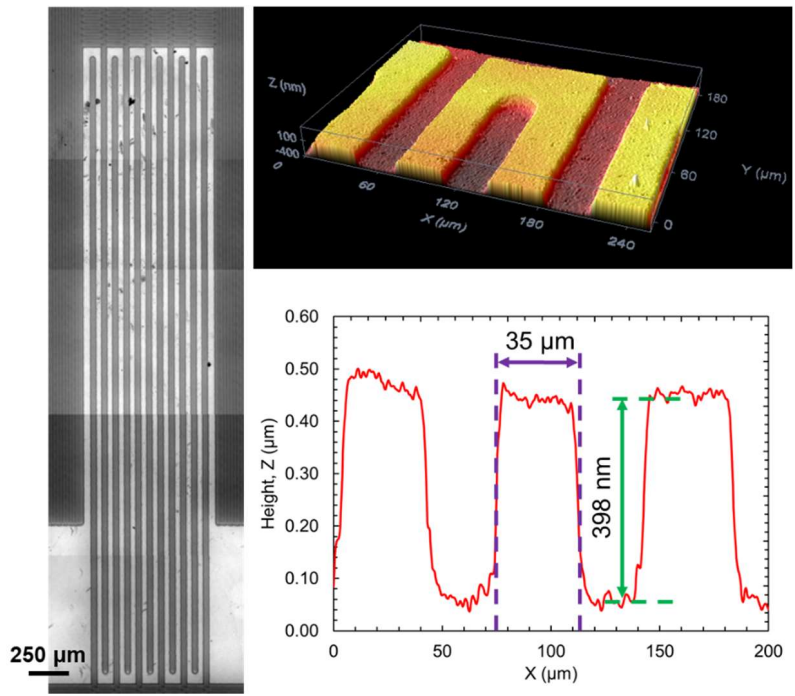

*Figure 11. Photograph (left), 3D topography (right top) and corresponding 1D cross-sectional profile (right bottom) in a circuit with 12 paths machined with the fs-UV (343 nm) laser wavelength.*

These results evidence that this technology has a great potential in the industrial realm. It has been demonstrated that, in addition to precise control over all variables of laser micromachining, scalability can be achieved, leveraging industrial manufacturing processes such as batch production or processing of large surfaces. This represents a starting point for implementing thin film micromachining of transparent conductive oxide thin films or even conducting polymers on an industrial scale, providing a competitive advantage over more traditional techniques.

## 4.- Conclusions

This study provides a comprehensive analysis of the impact of laser-induced periodic surface structures (LIPSS) on the electrical properties of laser micromachined 115 - 400 nm ITO thin films. Our study highlights the distinct effects of laser wavelength on LIPSS morphology. UV (343 nm) and green (515 nm) lasers produce markedly different structures, owing to variations in energy deposition with the ITO film and intra-pulse scattering/interference effects. Thus, UV laser radiation creates sharper lateral transitions between the pristine and machined regions, reducing the zones of accompanying LIPSS, whereas green laser radiation results in smoother transitions and wider nanotextured regions. These differences have been explained through the formation of different type of submicrometric LIPSS (HSFL-I and LSFL-I) ruled by the wavelength of the ultrashort laser pulses. Moreover, no significant changes in the overall chemical composition of the ITO films were observed between pristine and the laser nanostructured (percolative) regions at the edges, maintaining a constant Sn/In ≈ 0.09 atomic ratio.

By measuring the electrical resistance in four series of ITO tracks of variable width, machined using fs laser subtractive manufacturing method we have demonstrated that LIPSS orientation and periodicity play an important role in the circuit electrical properties. Each series was machined using UV or green laser with polarization either along or across the ITO tracks. When using the green laser wavelength, the orientation of the LIPSS markedly influences the electrical behavior: circuits with LIPSS oriented perpendicular to the electrical lines exhibit greater resistivity than those with parallel orientation. Using a simple electrical resistance model, we have estimated that the resistance increases by a factor ≈ 1.55 and ≈ 3.5 compared to pristine film, for LIPSS oriented parallel or perpendicular to the path, respectively (for given length and width dimensions). This effect becomes particularly significant for track widths narrower than 80 μm. In contrast, when machining with the UV wavelength, such anisotropy could not be accurately quantified. In the latter, even for narrow conductive ITO lines ($w_0$ ≈ 20–25 μm), the contribution of the laser-affected regions to the total resistance can be neglected within about 5 % accuracy. These results underscore the importance of tailoring LIPSS characteristics to optimize electrical conductivity in laser micromachined circuits.

Overall, these findings establish laser micromachining as a precise and scalable method for laser-fabricating of thin-film based microelectronic circuits. The ability to control LIPSS morphology and distribution offers significant potential for tailoring electrical properties to meet specific application requirements. Future work should focus on integrating these optimized circuits into functional devices, exploring their application in fields such as sensors, displays, and photovoltaics. Additionally, advancements in laser technology and parameter optimization could further enhance the precision and efficiency of this micromachining approach.

**Data availability**

Data are available at: https://doi.org/10.5281/zenodo.17206943.

**Acknowledgments**

The authors are grateful for funding from the European Commission EU H2020 program under grant agreement 899352 (FETOPEN-01-2018-2019-2020 – SOUNDofICE), from the Spanish research projects PID2020-113034RB-I00 (funded by MCIN/AEI/10.13039/501100011033), PID2023-146041OB-C21 (funded by MICIU/AEI/10.13039/501100011033 and ERDF/EU) and TED2021-130916B-I00 (funded by MICIU/AEI/10.13039/501100011033, ERDF (FEDER)" A way of making Europe, Fondos Nextgeneration EU and Plan de Recuperación, Transformación y Resiliencia) and from the Government of Aragon (Research Group T54_23R). Authors also thank the use of Servicio General de Apoyo a la Investigación – SAI and the Spanish National Facility ELECMI ICTS, node "Laboratorio de Microscopías Avanzadas (LMA)" at Universidad de Zaragoza and Servicio de Microscopía Electrónica de Barrido de Alta Resolución at Universidad Pablo de Olavide.

**References**
1. Alam, MJ, Cameron DC. Optical and electrical properties of transparent conductive ITO thin films deposited by sol-gel process, *Thin Solid Films*, **2000**, 377-378, 455-459. https://doi.org/10.1016/S0040-6090(00)01369-9
2. Singh G, Sheokand H, Ghosh S, Srivastava KV, Ramkumar J, Ramakrishna SA. Excimer laser micromachining of indium tin oxide for fabrication of optically transparent metamaterial absorbers. *Appl Phys A* **2019**, 125, 23. https://doi.org/10.1007/s00339-018-2013-7
3. Chou TR, Chen SH, Chiang YT, Lin YT, Chao CY. Highly conductive PEDOT: PSS films by post-treatment with dimethyl sulfoxide for ITO-free liquid crystal display. J *Mater Chem C* **2015**, 3, 3760–3766. https://doi.org/10.1039/C5TC00276A
4. Minami T. Present status of transparent conducting oxide thin-film development for Indium-Tin-Oxide (ITO) substitutes. *Thin Solid Films* **2008**, 516, 5822–5828. https://doi.org/10.1016/j.tsf.2007.10.063
5. Park M, Chon BH, Kim HS, Jeoung SC, Kim D, Lee JI, Chu HY, Kim HR. Ultrafast laser ablation of indium tin oxide thin films for organic light-emitting diode application. *Opt. Lasers Eng.* **2006**, 44, 138–146.

https://doi.org/10.1016/j.optlaseeng.2005.03.009

6. Kim JS, Friend RH, Cacialli F. Improved operational stability of polyfluorene-based organic light-emitting diodes with plasma-treated indium–tin–oxide anodes. *Appl Phys Lett.* **1999**, 74, 3084–3086. https://doi.org/10.1063/1.124069

7. Bian Q, Yu X, Zhao B, Chang Z, Lei S. Femtosecond laser ablation of indium tin-oxide narrow grooves for thin film solar cells. *Opt. Laser Technol.* **2013**, 45, 395–401. https://doi.org/j.optlastec.2012.06.018

8. Cheng CW, Shen WC, Lin CY, Lee YJ, Chen JS. Fabrication of micro/nano crystalline ITO structures by femtosecond laser pulses. *Appl. Phys. A* **2010**, 101, 243–248. https://doi.org/10.1007/s00339-010-5810-1

9. Aydın EB, Sezgintürk MK. Indium tin oxide (ITO): A promising material in biosensing technology. *TrAC - Trends Anal. Chem.* **2017**, 97, 309–315. https://doi.org/10.1016/j.trac.2017.09.021

10. Molpeceres C, Lauzurica S, Ocaña JL, Gandía JJ, Urbina L, Cárabe J. Microprocessing of ITO and a-Si thin films using ns laser sources. *J Micromechanics Microengineering* **2005**, 15, 1271–1278. https://doi.org/10.1088/0960-1317/15/6/019

11. European Commission. Critical and strategic materials [Internet]. RMIS – Raw Materials Information System. 2023 [cited 2025 August 09]. Available from: https://rmis.jrc.ec.europa.eu/eu-critical-raw-materials

12. Choi HW, Farson DF, Bovatsek J, Arai A, Ashkenasi D. Direct-write patterning of indium-tin-oxide film by high pulse repetition frequency femtosecond laser ablation. *Appl Opt.* **2007**, 46, 5792–5799. https://doi.org/10.1364/ao.46.005792

13. Krause S, Miclea PT, Steudel F, Schweizer S, Seifert G. Precise microstructuring of indium-tin oxide thin films on glass by selective femtosecond laser ablation. *EPJ Photovoltaics* **2013**, 4, 40601. https://doi.org/10.1051/epjpv/2012013

14. Altinkaya C, Najmi MA, Lida D, Ohkawa K. Light extraction improvement via ITO p-electrodes for InGaN red micro-LEDs emitting at 640 nm. *Opt Contin.* **2025**, 4, 1040–1050. https://doi.org/10.1364/OPTCON.559350

15. Datta RS, Syed N, Zavabeti A, Jannat A, Mohiuddin M, Rokunuzzaman M, Zhang BY, Rahman MA, Atkin P, Messalea KA, Ghasemian MB, Gaspera ED, Bhattacharyya S, Fuhrer MS, Russo SP, McConville CF, Esrafilzadeh D, Kalantar-Zadeh K, Daeneke T, Flexible two dimensional indium tin oxide fabricated using a liquid metal printing technique, *Nature Electonics* **2020**, 3, 51-58. https://doi.org/10.1038/s41928-019-0353-8

16. Nsabimana J, Wang Y, Ruan Q, Li T, Shen H, Yang C, Zhu Z, An electrochemical method for a rapid and sensitive immunoassay on digital microfluidics with integrated indium tin oxide electrodes coated on a PET film, *Analyst* **2021**, 146, 4473-4479. https://doi.org/10.1039/d1an00513h

17. Papanastasiou DT, Sekkat A, Nguyen VH, Jiménez C, Muñoz-Rojas D, Bruckert F, Bellet D, Stable Flexible Transparent Electrodes for Localized Heating of Lab-on-a-chip Devices, *Adv. Mater. Technol.* **2023**, 8, 2200563. https://doi.org/10.1002/admt.202200563

18. Park JH, Seok HJ, Jung SH, Cho HK, Kim HK, Rapid thermal annealing effect of transparent ITO source and drain electrode for transparent thin film transistors, *Ceramics International* **2021**, 47, 3149-3158. https://doi.org/10.1016/j.ceramint.2020.09.152

19. Wang Z, Huang Z, Lu N, Guan J, Hu Y. Energy transfer and patterning characteristics in pulsed-laser subtractive manufacturing of single layer of $MoS_2$. *Int J Heat Mass Transf.* **2023**, 204, 123873. https://doi.org/10.1016/j.ijheatmasstransfer.2023.123873.

20. Zheng C, Cai Y, Zhang P, Zhang T, Aslam J, Song Q, Liu Z, Femtosecond laser precision machining of carbon film based on aramid paper substrate. *J Manuf Process.* **2024**, 119, 57–65. https://doi.org/10.1016/j.jmapro.2024.03.061

21. Liu X, Lu Z, Jia Z, Chen Z, Wang X. Sandwich-structured ZnO-$MnO_2$-ZnO thin film varistors prepared via magnetron sputtering. *J Eur Ceram Soc.* **2023**, 43, 3344–3350. https://doi.org/10.1016/j.jeurceramsoc.2023.01.030
22. Cheng CW, Lin CY. High precision patterning of ITO using femtosecond laser annealing process. *Appl Surf Sci*. **2014**, 314, 215–220. https://doi.org/10.1016/j.apsusc.2014.06.174
23. Bonse J, Wrobel JM, Brzezinka KW, Esser N, Kautek W. Femtosecond laser irradiation of indium phosphide in air: Raman spectroscopic and atomic force microscopic investigations. *Appl Surf Sci.* **2002**, 202, 272–282. https://doi.org/10.1016/S0169-4332(02)00948-0
24. Lenzner M, Krüger J, Kautek W, Krausz F. Precision laser ablation of dielectrics in the 10 fs regime. *Appl. Phys. A*, **1999,** 68, 369–371. https://doi.org/10.1007/s003390050906
25. Bovatsek J, Tamhankar A, Patel RS, Bulgakova NM, Bonse J. Thin film removal mechanisms in ns-laser processing of photovoltaic materials. *Thin Solid Films*, **2010**, 518, 2897–2904. https://doi.org/10.1016/j.tsf.2009.10.135
26. Porta-Velilla L, Martínez E, Frechilla A, Castro M, de la Fuente GF, Bonse J, Angurel LA. Grain Orientation, Angle of Incidence, and Beam Polarization Effects on Ultraviolet 300 ps-Laser-Induced Nanostructures on 316L Stainless Steel. *Laser Photonics Rev.* **2024**, 18, 2300589. https://doi.org/10.1002/lpor.202300589
27. Jiang Q, Zhang Y, Xu Y, Zhang S, Feng D, Jia T, Sun Z, Qiu J. Extremely High-Quality Periodic Structures on ITO Film Efficiently Fabricated by Femtosecond Pulse Train Output from a Frequency-Doubled Fabry – Perot Cavity. *Nanomaterials*, **2023**, 13, 1510, https://doi.org/10.3390/nano13091510
28. Bonse J, Kirner SV, Krüger J, Laser-Induced Periodic Surface Structures (LIPSS). In: Sugioka, K. (eds) Handbook of Laser Micro- and Nano-Engineering. Springer, Cham, 2021 https://doi.org/10.1007/978-3-030-63647-0_17
29. Bonse J. Quo vadis LIPSS?—recent and future trends on laser-induced periodic surface structures. *Nanomaterials*, **2020**, 10, 1950, https://doi.org/10.3390/nano10101950
30. Lopez-Santos C, Puerto D, Siegel J, Macias-Montero M, Florian C, Gil-Rostra J, López-Flores V. Borras A, González-Elipe AR, Solís J, Anisotropic Resistivity Surfaces Produced in ITO Films by Laser-Induced Nanoscale Self-organization. *Adv Opt Mater.* **2021**, 9, 2001086. https://doi.org/10.1002/adom.202001086
31. Wonneberger R, Gräf S, Bonse J, Wisniewski W, Freiberg K, Hafermann M, Ronning C, Müller FA, Undisz A, Tracing the Formation of Femtosecond Laser-Induced Periodic Surface Structures (LIPSS) by Implanted Markers. *ACS Appl. Mater. Interfaces*, **2025**, 17, 2462–2468. https://doi.org/10.1021/acsami.4c14777
32. Porta-Velilla L, Turan N, Cubero Á, Shao W, Li H, De la Fuente GF, Martínez E, Larrea A, Castro M, Koralay H, Cavdar S, Bonse J, Angurel LA. Highly Regular Hexagonally-Arranged Nanostructures on Ni-W Alloy Tapes upon Irradiation with Ultrashort UV Laser Pulses, *Nanomaterials*, **2022**, 14, 2380 https://doi.org/10.3390/nano12142380
33. Bonse J, Krüger J, Höhm S, Rosenfeld A. Femtosecond laser-induced periodic surface structures. *J Laser Appl*. **2012**, 24, 042006. https://doi.org/10.2351/1.4712658
34. Frechilla A, Sekkat A, Dibenedetto M, lo Presti F, Porta-Velilla L, Martínez E, De la Fuente GF, Angurel LA, Muñoz-Rojas D, Generating colours through a novel approach based on spatial ALD and laser processing. *Mater Today Adv.* **2023**, 19, 100414, https://doi.org/10.1016/j.mtadv.2023.100414
35. Bonse J, Krüger J. Structuring of thin films by ultrashort laser pulses. *Appl Phys A.* **2023**, 129, 14, https://doi.org/10.1007/s00339-022-06229-x
36. Liu JM. Simple technique for measurements of pulsed Gaussian-beam spot sizes. Opt Lett. 1982, 7, 196-198. https://doi.org/10.1364/OL.7.000196

37. Bonse, J, Gräf, S. Maxwell Meets Marangoni—A Review of Theories on Laser-Induced Periodic Surface Structures. *Laser & Photonics Reviews* **2020**, 14, 2000215. https://doi.org/10.1002/lpor.202000215
38. Rudenko A, Colombier J-P, Höhm S, Rosenfeld A, Krüger J, Bonse J, Itina TE. Spontaneous periodic ordering on the surface and in the bulk of dielectrics irradiated by ultrafast laser: a shared electromagnetic origin. *Sci. Rep.* **2017**, 7, 12306. https://doi.org/10.1038/s41598-017-12502-4
39. Cubero Á, Martínez E, Angurel LA, de la Fuente GF, Navarro R, Legall H, FrÜger J, Bonse J, Surface superconductivity changes of niobium sheets by femtosecond laser-induced periodic nanostructures. *Nanomaterials*. **2020**, 10, 25251–16. https://doi.org/10.3390/nano10122525
40. Pan A, Dias A, Gomez-Aranzadi M, Olaizola SM, Rodriguez A. Formation of laser-induced periodic surface structures on niobium by femtosecond laser irradiation. *J Appl Phys.* **2014**, 115, 173101. https://doi.org/10.1063/1.4873459

# Influence of edge Laser-Induced Periodic Surface Structures (LIPSS) on the electrical properties of fs laser-machined ITO microcircuits

A. Frechilla[1], E. Martínez[1], J. del Moral[2], C. López-Santos[2,3], J. Frechilla[1], F. Nuñez-Gálvez[2,3], V. López-Flores[2,3], G.F. de la Fuente[1], D. Hülagü[4], J. Bonse[4], A. R. González-Elipe[2], A. Borrás[2], L.A. Angurel[1]

1. Instituto de Nanociencia y Materiales de Aragón, INMA, CSIC-Universidad de Zaragoza, María de Luna, 3, 50018 Zaragoza, Spain

2. Nanotechnology on Surfaces and Plasma, Instituto de Ciencia Materiales de Sevilla, ICMS, CSIC-Universidad de Sevilla, Américo Vespucio 49, 41092 Sevilla, Spain

3. Dpto. Física Aplicada I. Escuela Politécnica Superior. Universidad de Sevilla. c/ Virgen de África 7, 41011 Sevilla, Spain

4. Bundesanstalt für Materialforschung und –prüfung (BAM), Unter den Eichen 87, 12205 Berlin, Germany

**High-resolution electrical measurements with the 4-point microprobe station**

High-resolution measurements were performed in a 4-point probe configuration with a contact spacing of 60 µm, as illustrated in Figure S1.

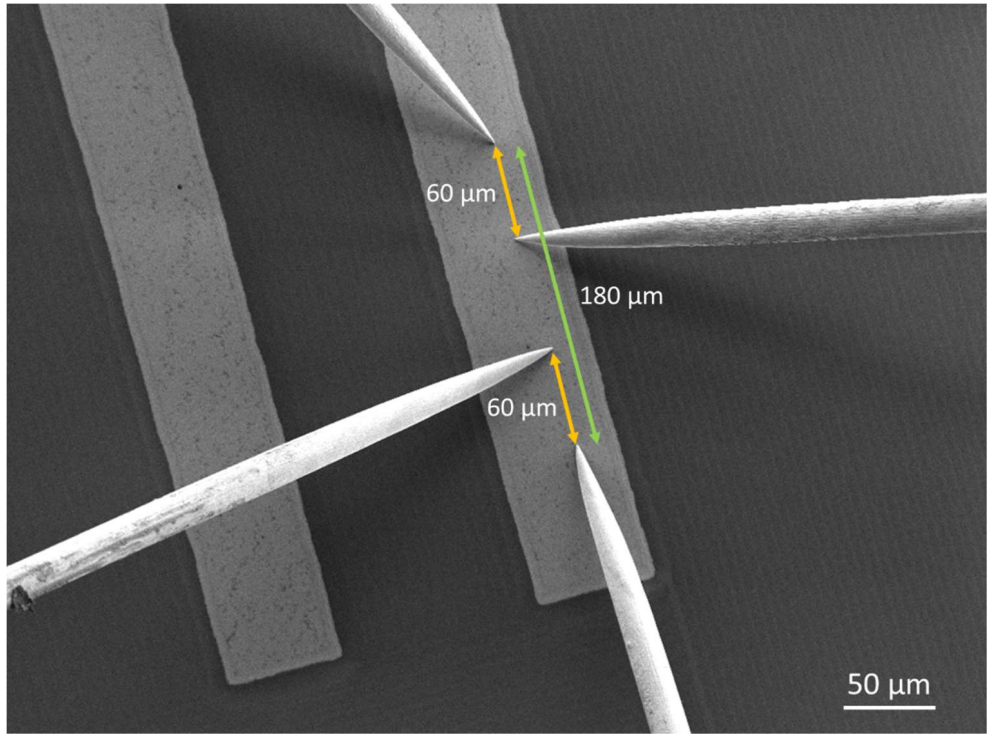

***Figure S1.*** *High-resolution FESEM image of the 4-point probe measurement setup using a Kleindiek micromanipulator system inside a Zeiss GeminiSEM 300. The distance between the voltage contacts was set to 60 µm for microscale electrical characterization, as indicated coloured by orange double-arrows.*

**Study of the LIPSS spatial periodicity evolution close to the machined region**

Figure S2 shows the regions where the LIPSS spatial periodicity has been analyzed in the sample machined with the λ = 515 nm laser and some examples of the 2D-FFT profiles used to quantify it.

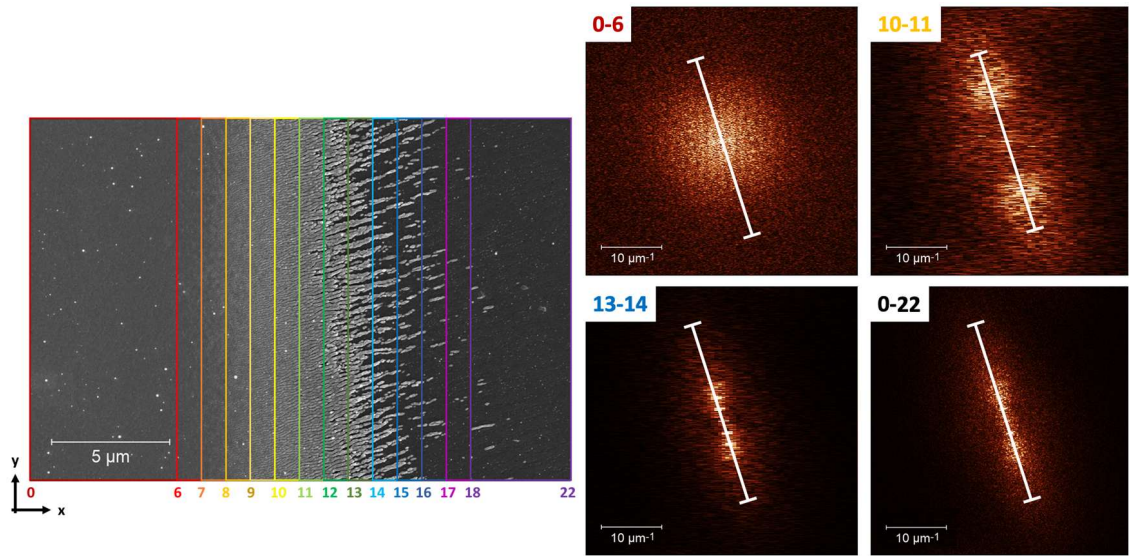

*Figure S2. Analysis of the spatial periodicity evolution along the ITO/glass boundary for λ = 515 nm. At the left side of the image: Top-view of the complete SEM image showing the position ranges of the x-coordinate (in μm) that define the different sections analysed by 2D-FFT. The right side of the figure displays some 2D-FFT images showing the lines for the 1D-profiles that are used to quantify the LIPSS spatial periodicity across these transitional zones.*

**Microprobe WDS measurements**

Figure S3 presents the raw data obtained from Microprobe WDS measurements at the edge of the ITO path micromachined using a green laser with a wavelength of 515 nm.

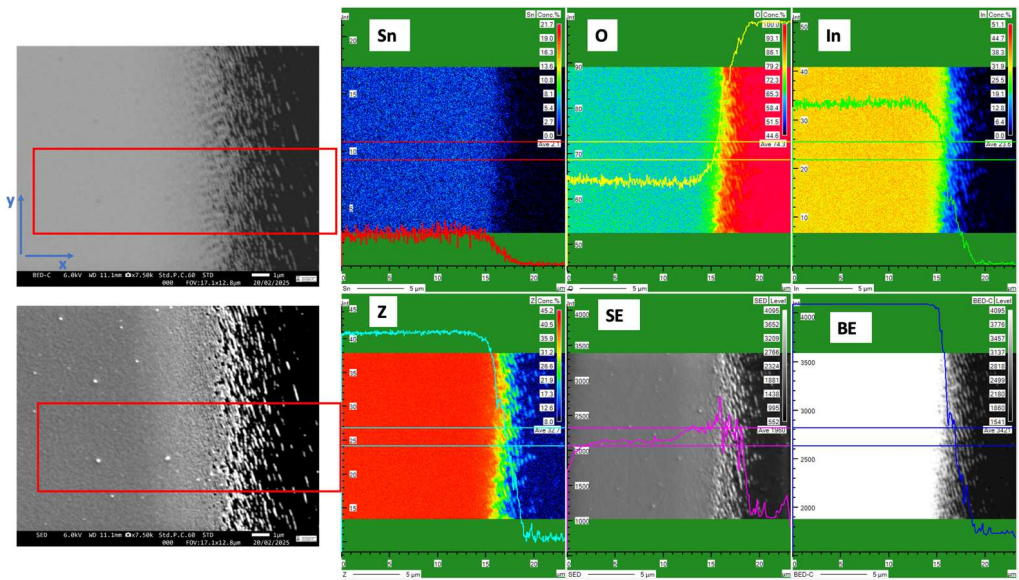

***Figure S3.*** *Microprobe WDS measurements (right panels) of the edge of the ITO path micromachined used a green laser, with the LIPSS parallel to the path. The original ITO surface appears in the left part of the scanning electron micrographs (left column of panels), and the glass substrate on the right one. The profiles correspond to the shown area (Δy = 5 μm). Left panels correspond to SEM views of the analysed surface for backscattered (top) and secondary (bottom) electron detectors.*

Figure S4 displays the corresponding raw data for the ITO path edge micromachined using a UV laser with a wavelength of 343 nm.

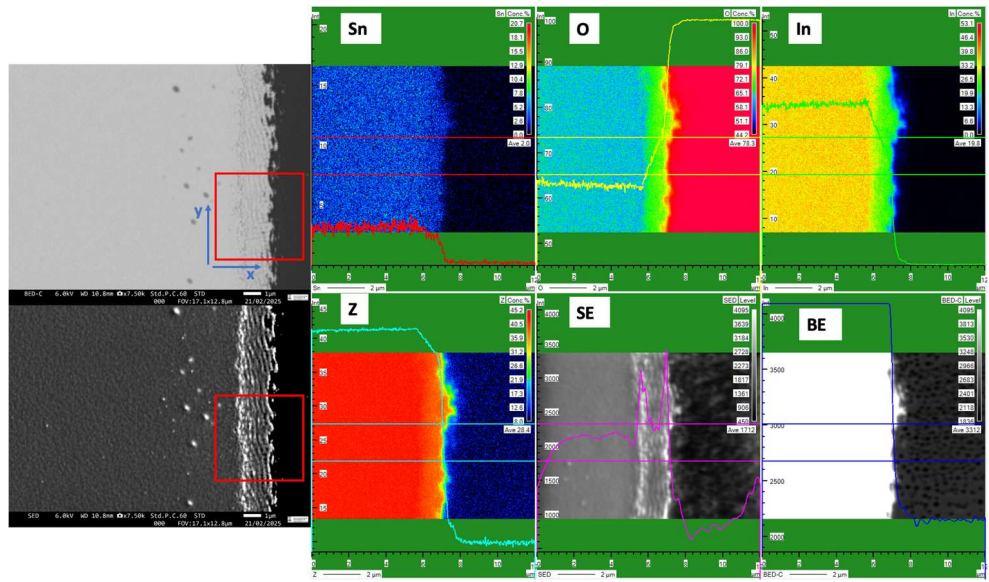

***Figure S4.*** *Microprobe WDS measurements of the edge of the ITO path micromachined used an UV laser, with the LIPSS parallel to the path. The original ITO surface appears in the left part of the photographs, and the glass substrate on the right one. The profiles correspond to the shown area (Δy = 5 μm). Left panels correspond to SEM views of the analysed surface for backscattered (top) and secondary (bottom) electron detectors*

**Electrical characterization**

A top-view photograph of a 12-track set, each designed with its corresponding current and voltage contacts for four-point resistance measurements, can be seen in Figure S5.

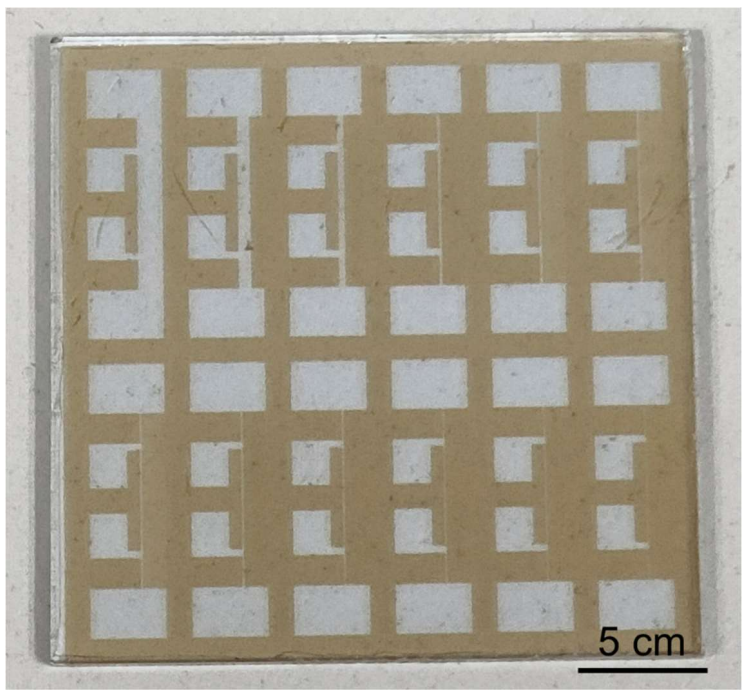

*Figure S5. Photograph of a 12-track set sample designed with the corresponding current and voltage contacts for four-point resistance measurements.*